\documentclass{ifacconf}

\usepackage{graphicx}      
\usepackage{bm}
\usepackage{amsmath,amssymb,amsfonts}
\usepackage{textcomp}
\usepackage{xcolor}
\usepackage{comment}
\usepackage{optidef}
\usepackage[T1]{fontenc}
\usepackage{mathtools} 
\usepackage{cuted}
\usepackage{flushend}
\usepackage{bbm}
\usepackage{dsfont}
\usepackage{mathtools}
\usepackage{multirow}
\usepackage{float}
\usepackage{dirtytalk}
\usepackage{array}

\usepackage{multirow}
\usepackage{booktabs}
\usepackage{subcaption}
\usepackage{siunitx}
\usepackage{array}

\usepackage{natbib}        
\begin{document}
\begin{frontmatter}

\title{A Deep-Unfolding Approach to RIS Phase Shift Optimization Via Transformer-Based Channel Prediction
} 

\thanks[footnoteinfo]{The research leading to this paper was supported by the Research Council of Finland (former Academy of Finland) {6G Flagship program} (Grant Number: 346208), and Business Finland's 6GBridge program through the projects Local 6G (Grant Number 8002/31/2022) and 6CORE (Grant Number 8410/31/2022).}

\author{Ishan Koralege} ,
\author{Arthur S. de Sena}, 
\author{Nurul H. Mahmood},
\author{Farjam Karim},
\author{Dimuthu Lesthuruge},
\author{Samitha Gunarathne}

\address{Centre for Wireless Communications, University of Oulu, \\Oulu, Finland } 
\{ishan.koralege, arthur.sena, nurulhuda.mahmood,  farjam.karim, lesthuruge.silva, samitha.gunarathne\}@oulu.fi

\begin{abstract}                
Reconfigurable intelligent surfaces (RISs) have emerged as a promising solution that can provide dynamic control over the propagation of electromagnetic waves. The RIS technology is envisioned as a key enabler of sixth-generation networks by offering the ability to adaptively manipulate signal propagation through the smart configuration of its phase shift coefficients, thereby optimizing signal strength, coverage, and capacity. However, the realization of this technology's full potential hinges on the accurate acquisition of channel state information (CSI). In this paper, we propose an efficient CSI prediction framework for a RIS-assisted communication system based on the machine learning (ML) transformer architecture. Architectural modifications are introduced to the vanilla transformer for multivariate time series forecasting to achieve high prediction accuracy. The predicted channel coefficients are then used to optimize the RIS phase shifts. Simulation results present a comprehensive analysis of key performance metrics, including data rate and outage probability. Our results confirm the effectiveness of the proposed ML approach and demonstrate its superiority over other baseline ML-based CSI prediction schemes such as conventional deep neural networks and long short-term memory architectures, albeit at the cost of slightly increased complexity.
\end{abstract}

\begin{keyword}
Channel prediction, deep neural network, machine learning, reconfigurable intelligent surface, transformer.
\end{keyword}

\end{frontmatter}

\section{Introduction}
Sixth-generation (6G) 
wireless networks will have the ability to push beyond the limits of today's wireless systems with their groundbreaking expected features such as very low latency, ultra-high reliable connectivity, enhanced data security, and integrated intelligence that leverages machine learning (ML) capabilities  \citep{IEEEJournal6GWireless_Chowdhury}. Within this framework, reconfigurable intelligent surface (RIS) has emerged as a potential game-changer. RIS introduces a new layer of intelligence into the physical layer, allowing the stochastic wireless propagation environment to be somewhat controlled \citep{mahmood23_functional}. This technology holds the promise of substantial improvements in signal strength, signal coverage, and network capacity by offering advanced control over the properties of electromagnetic waves \citep{wirelessRIS_Basar}. RISs are composed of passive reflecting elements that can be dynamically configured to manipulate the way wireless signals propagate toward receivers, enabling the achievement of diverse goals. 

Since RISs typically comprise only nearly passive components with no data processing capabilities, acquiring RIS-associated channel state information (CSI) is a fundamental challenge \citep{RIS_challenges_yuan}. Classical methods of CSI estimation such as least squares and minimum mean-squared error, as in the work in \citep{Classical_channel_est_Ardah}, rely on pilot signals, which incurs significant signaling overhead and channel acquisition delay, especially when the number of RIS elements is large. It is further exacerbated under dynamic wireless environments when the channel coherence time is short, thus requiring frequent CSI acquisition. These issues can be mitigated via learning-based approaches as they harness the power of ML to learn the optimal RIS phase shifts, thereby 
eliminating the requirement of complex mathematical modeling or overwhelming pilot training \citep{hashemi23_DRL}. However, such an approach suffers from slow convergence, large training overhead, and poor generalizability. A deep unfolding approach, where ML-based techniques are used to learn partial system blocks while adhering to conventional optimization approaches for the overall system design \citep{Stimming19_deepUnfolding} can be adopted to address these limitations. This results in improved performance at a much lower complexity. More specifically, we propose to replace the pilot-based CSI estimation procedure with an ML-based CSI prediction method, which is then used to mathematically optimize the RIS phase shifts.

ML-assisted solutions have the potential to adapt and learn the dynamics of the CSI autonomously. By training ML models with large data sets, unforeseen channel characteristics can be captured, even in complex communications environments assisted by RISs. Conventional deep neural network (DNN) and recurrent neural network (RNN) architectures can provide, to some extent, satisfactory prediction results \citep{DNNEst}. However, both DNNs and RNNs have limitations such as the vanishing gradient phenomenon. This occurs when training DNNs with a large number of layers, including activation functions. In such cases, the gradients used to update the network become extremely small or even vanish as they are backpropagated. As a result, the convergence of the algorithm is substantially slowed. \citep{vanishing}. Long short-term memory (LSTM) networks are an evolution of RNNs that have been proposed to prevent the vanishing gradient problem \citep{LSTMvanish}, which enables the processing of longer data sequences. A recent 
architecture named \emph{transformers} pushes the boundaries further. Transformers have excelled across various domains due to their superior characteristic performance \citep{Efficient_Tx_Tay}, significantly outperforming most of the previous deep learning approaches. 

Diverse ML-based strategies have been employed in the literature to predict RIS-associated CSI. In \citep{federated_est}, a federated learning strategy with distributed convolution neural networks (CNNs) was used for channel prediction in a RIS-assisted multi-antenna system employing orthogonal frequency-division multiplexing. A distributed CNN framework for downlink channel prediction in a narrow-band multi-user system was proposed in \citep{DML_NN}. A real-time reinforcement learning-aided CSI measurement scheme for a RIS-assisted millimeter wave (mmWave) system was considered in \citep{dis_RL}. The authors in \citep{DL_FC} developed schemes based on CNNs and fully connected DNNs for RIS channel extrapolation and beam searching in a single-antenna system. A deep denoising CNN was used for aiding RIS channel estimation in a mmWave system in \citep{ddnn}. In \citep{CNN_LSTM}, a strategy based on CNNs and LSTMs was proposed to predict the CSI of a RIS-assisted system employing non-orthogonal multiple access, and in \citep{Trans_Parallel}, a transformer-aided scheme was proposed for predicting the CSI of an uplink RIS-assisted mmWave system.

To the best of our knowledge, only the work in \citep{Trans_Parallel} has exploited the transformer architecture for channel prediction in RIS-assisted networks. In this paper, we propose a novel transformer-based approach to predict the unknown CSI in a RIS-assisted communication system. The predicted CSI is then used to optimize the RIS phase shifts for downlink data transmission. We compare our proposed prediction scheme with other state-of-the-art learning-based approaches, evaluating the performance using metrics such as data rate and outage probability. Based on the ML architectures we developed, it is shown that the transformer outperforms LSTM and DNN architectures, though the architecture is slightly more complex. Furthermore, the DNN architecture shows the lowest performance with the least complexity, while the LSTM method demonstrates higher performance and greater complexity than the DNN approach, yet lower performance and less complexity compared to the transformer architecture.


The remainder of the article is organized as follows. In Section \ref{sec:systemModel}, the system model is provided. Section \ref{sec:CSI Prediction and Data Preparation} gives insights into the proposed ML-based CSI predictor. Section \ref{sec:results} provides the results and discussion. Finally, the conclusions are presented in Section \ref{sec:conc}.


\section{System Model}
\label{sec:systemModel}

\begin{figure}[tb]
    \centering
    \includegraphics*[width=0.5\textwidth]{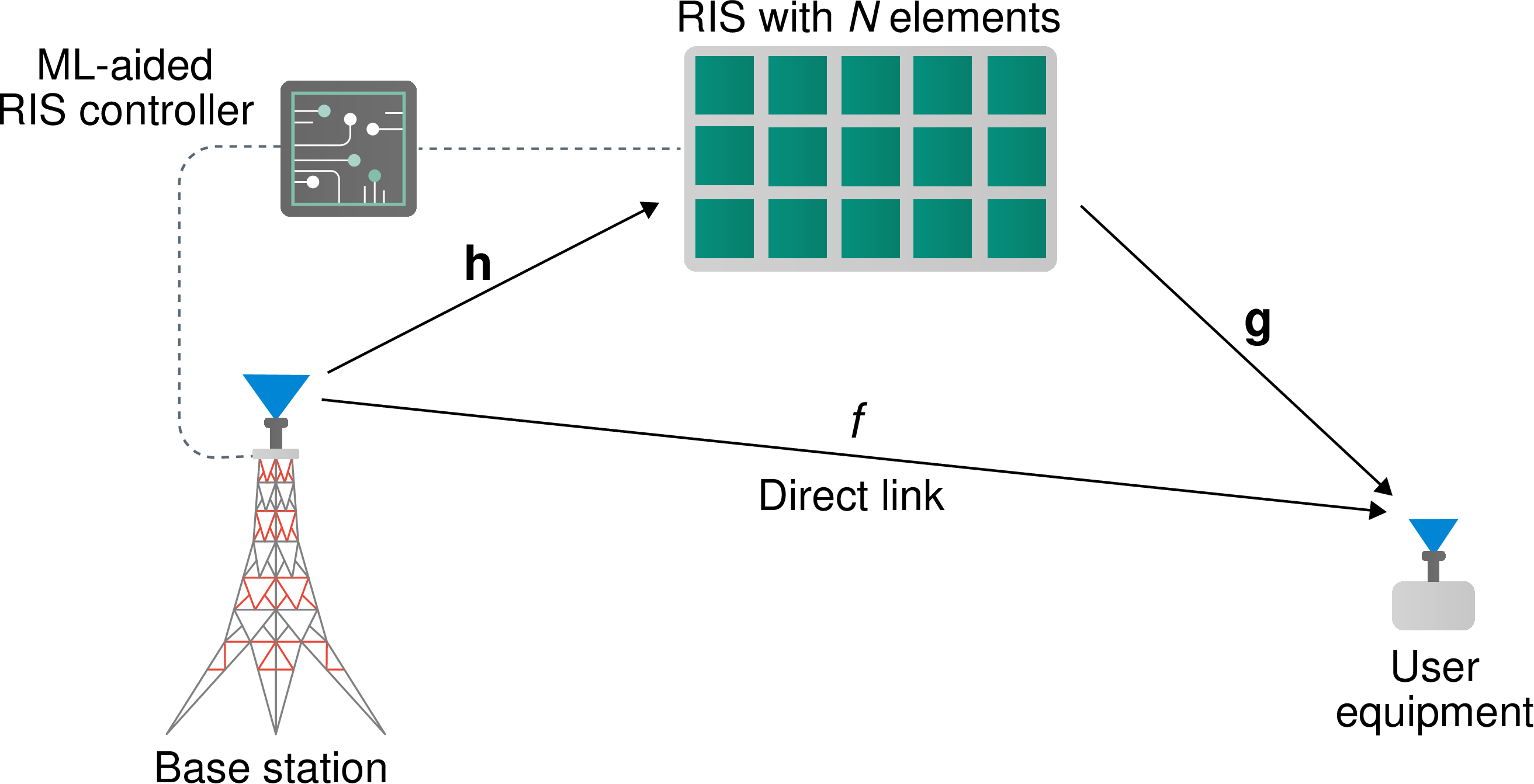}
    \caption{System model. An RIS comprising $N$ elements assists the communication between one single-antenna BS and one single-antenna UE.}
    \label{fig:system_model}
\end{figure}

Consider a downlink RIS-aided single-input single-output (SISO) communication system as in Fig. \ref{fig:system_model}, where the base station (BS) and the user equipment (UE) are each equipped with a single antenna, and the RIS comprises $N$ reflecting elements. Let $f \in \mathbb{C}$, $\mathbf{h} \in \mathbb{C}^{N \times 1}$, $\mathbf{g}\in \mathbb{C}^{N \times 1}$ be the fading channels between the BS and the UE, the BS and the RIS, and the RIS and the UE, respectively. The channel vectors $\mathbf{h} = \left[ h_1, \dots, h_i, \dots, h_{N} \right]^T$ and $\mathbf{g} = \left[ g_1, \dots, g_i, \dots, g_{N} \right]^T$, where $h_i$ $(g_i)$ is the channel between the BS (UE) and $i^{th}$ RIS element. We assume that the channel coefficients are Rayleigh distributed, i.e., $f \sim \mathcal{CN}(0,\,L_{\text{\tiny BS-UE}})$, $h_i \sim \mathcal{CN}(0,\,L_{\text{\tiny BS-RIS}})$, and $g_i  \sim \mathcal{CN}(0,\,L_{\text{\tiny RIS-UE}})$, for $i= 1 \cdots, N$, where $L_a$ denotes the path loss coefficients in linear scale for the $a^{th}$ link with $a \in\{\text{BS-UE}, \text{BS-RIS}, \text{RIS-UE}\}$. In this work, we consider the log-distance path loss model, which can be given in dB by 
\begin{equation}
\label{eqn:path loss}
L_{a, \text{dB}} = L_{0, \text{dB}} + 10\eta_a\log_{10}\left(\frac{d_a}{d_0}\right),
\end{equation}
where $\eta_a$ and $d_a$ represent the path loss exponent and the distance for the link $a$ respectively, and $L_{0, \text{dB}}$ is the reference path loss in dB at a reference distance $d_0$. Under this model, the received signal at the UE can be expressed as
 \begin{align}
 \label{eqn:Received_Signal}
 y =  \sqrt{P}\left(f + \mathbf{g}^{T}\mathbf{\Phi}\mathbf{h}\right)x + n ,
\end{align}
where $n \sim \mathcal{CN}(0, N_0)$ is the additive white Gaussian noise (AWGN), $x$ is the transmitted symbol, $P$ represents the transmit power and $\bm{\Phi}$ is the matrix comprising the reflection coefficients of the RIS, which is a diagonal matrix defined as \citep{wirelessRIS_Basar}
\begin{align} \label{eqn:phi}
      \mathbf{\Phi} = \mathrm{diag}(\pmb{\phi}),
\end{align}
where $\pmb{\phi} = \left[ a_1e^{j\phi_1}, a_2e^{j\phi_2}, \dots, a_Ne^{j\phi_{N}} \right]^T$ is the RIS phase shift vector, in which $a_i$ is the amplitude and, $\phi_i$ is the adjustable phase induced by the the $i$\textsuperscript{th} reflecting element. Due to the passive nature of the RIS, we can assume $a_i = 1, \forall i = 1, \cdots, N$.

\subsection{RIS Phase Shift Optimization}
We wish to compute a phase shift matrix $\mathbf{\Phi}$ that can maximize the signal power propagating through the RIS at the UE during downlink data transmission. 
To this end, we assume that the direct link is non-dominant. This can be the case when it is non-line of sight, while the BS-RIS and RIS-UE links are in line of sight. 
Under this assumption, we can recall the Cauchy-Schwartz inequality and achieved \citep{bjornson2024} 
\begin{align} \label{eqn:inequality}
|\mathbf{g}^{T}\mathbf{\Phi}\mathbf{h}|^2 = |(\mathbf{h} \odot \mathbf{g})^{T} \pmb{\phi}|^2  \leq \left( \sum_{i=1}^{N}|h_i g_i e^{j\phi_i} | \right)^2,
\end{align}
where $\odot$ denotes the Hadamard product. The inequality in \eqref{eqn:inequality} provides an upper bound to the reflected power i.e., the maximum power that can be reflected by the RIS. The equality can be achieved if and only if $\phi_i = - \mathrm{arg} (h_i g_i)$. As a result, the desired optimal RIS phase shift vector can be given by
\begin{align} \label{eqn:phiopt}
\pmb{\phi}^* = e^{- j \mathrm{arg} (\mathbf{h} \odot \mathbf{g})}.
\end{align}
The optimum phase shift coefficients in \eqref{eqn:phiopt} can be readily computed after predicting the CSI through ML, as discussed in the subsequent sections.

\begin{figure*}[ht]
    \centering
    \includegraphics*[width= 1.0\textwidth]{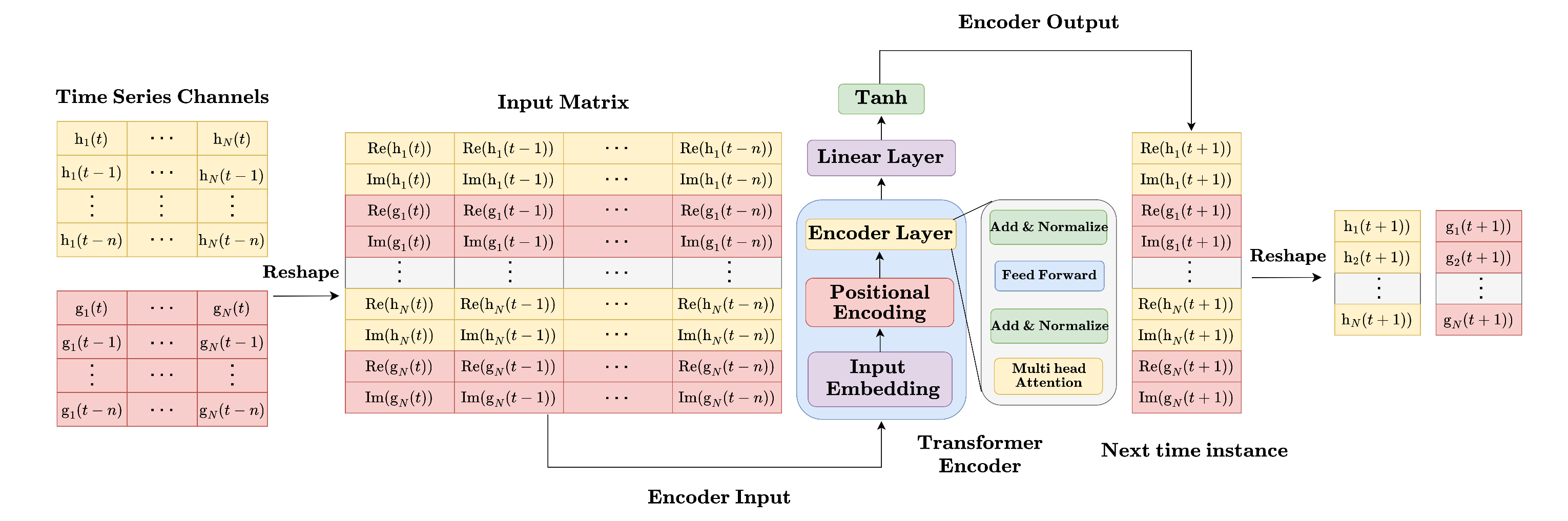}
    \caption{Sequence-to-one transformer architecture with input and output.}
    \label{fig:Tx_archi__model}
\end{figure*}

\subsection{Performance Metrics}
In this subsection, we present the performance metrics used to assess the effectiveness of the proposed ML-based CSI prediction framework. To this end, we first provide an expression for the signal-to-noise ratio (SNR). More specifically, by using  \eqref{eqn:Received_Signal} and \eqref{eqn:inequality}, the instantaneous SNR observed at the UE, denoted as $\gamma$, can be computed by
\begin{equation}
\label{eqn:gamma}
\gamma = \frac{\left|(\mathbf{h} \odot \mathbf{g})^{T} \pmb{\phi} + f\right|^2 P}{N_0}.
\end{equation}

The outage probability, $P_{\text{out}}$, is the first important performance metric used to assess the performance of the UE. It is the probability of the instantaneous SNR falling below a defined threshold $\gamma_{\text{th}}$, i.e., 
\begin{equation}
\label{eqn:outageprob}
P_{\text{out}} = \mathrm{P}(\gamma < \gamma_{\text{th}}),
\end{equation}
where $\mathrm{P}$ denotes probability. 

Last, we consider the maximum data rate that can be achieved when the BS is transmitting at a fixed rate $R_{\text{th}}$, subject to a given outage probability $P_{\text{out}}$, which can be obtained as
\begin{equation}
\label{eqn:achievable data rate}
R = (1 - P_{\text{out}}) R_{\text{th}},
\end{equation}
where $R_{\text{th}}$ represents the data rate threshold given by
\begin{equation}
\label{eqn:achievable data rateth}
R_{\text{th}} = \log_2(1 + \gamma_{\text{th}}).
\end{equation}

\section{CSI Prediction and Data Preparation}
\label{sec:CSI Prediction and Data Preparation}

This section presents the proposed architecture of our transformer model which is used to make multivariate time series predictions of the channel coefficients. Details on the generation of the data sets used to train, validate, and test the implemented ML models are also presented. 

\subsection{Transformer-Based CSI Prediction}

In this subsection, we describe our proposed CSI prediction approach based on the ML transformer architecture, which we later exploit to optimize the RIS phase shifts. Transformers are a deep learning architecture initially introduced in \citep{AttentionIsAll_Vaswani} by a research group in Google. The key feature of this architecture is its state-of-the-art attention mechanism. The attention mechanism is a computational method that focuses on specific parts of the input data sequence, assigning varying degrees of importance to the different parts. This mechanism helps to inform the model where to pay attention when processing data. The vanilla transformer architecture consists of two parts, an encoder and a decoder, which consists of a sequence-to-sequence architecture. In this method, the model generates an output data sequence according to the given input sequence.

In this work, we aim to predict the CSI at the next time instance based on a given sequence of previous CSI samples. Therefore, the original sequence-to-sequence architecture is modified into a sequence-to-one architecture using only one encoder. 
More specifically, the encoder of our transformer model comprises an input embedding module, a positional encoding module, and a transformer encoder module, as shown in Fig. \ref{fig:Tx_archi__model}. The input embedding module transforms the dimension of the input data to the model dimension of the subsequent inner layer of the transformer. Since the transformer lacks a recurrent structure as in recurrent neural networks, it feeds the positional information to the output of the embedding layer separately. After that, using the acquired knowledge from the input sequence, an abstract representation is generated by the transformer encoder. The encoder consists of a multi-head attention block, layer normalization blocks, and a feedforward layer, as illustrated in Fig. \ref{fig:Tx_archi__model}. Then, the encoder output is sent through a fully connected (FC) linear layer and an activation function. Figure \ref{fig:Tx_archi__model} shows the modified sequence-to-one transformer architecture used for CSI prediction, where we feed samples of real and imaginary channel data separately to forecast the next time instance of the channel coefficients as output.

\begin{figure}[t!]
    \centering
    \includegraphics[width= 0.49\textwidth]{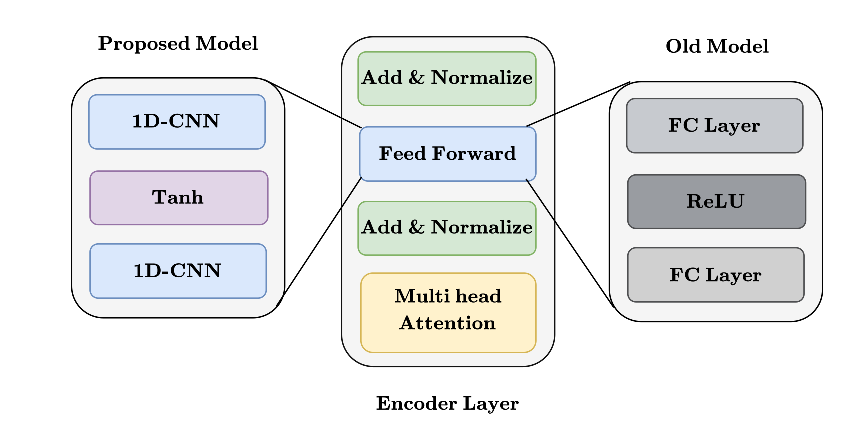}
    \caption{Feedforward layer in the vanilla transformer architecture (right) and the proposed model (left).}
    \label{fig:FCtoCnn}
\end{figure}  

The vanilla transformer architecture, originally introduced for natural language processing applications, contains two FC layers in the feedforward block and uses the rectified linear unit (ReLU) as the activation function \citep{AttentionIsAll_Vaswani}.
This setup is effective for capturing long-term variations in the multi-head attention output. However, given the relatively small window size, our problem requires observing and capturing short-term variations. Therefore, as shown in Fig. \ref{fig:FCtoCnn}, we modified the original architecture by implementing two one-dimensional (1D) CNN layers instead of the two FC layers, as CNNs effectively capture local temporal patterns crucial for accurate forecasting. Additionally, we replaced the ReLU activation function with a hyperbolic tangent (Tanh) function. These modifications result in better performance in our time series forecasting problem.

%

\subsection{Data Preparation}

Most of the cited works on CSI prediction assume independent/uncorrelated fading, whereas real-world scenarios often exhibit correlation. We incorporate this important property in our work by considering time-correlated Rayleigh-distributed fading channel coefficients. We first generate uncorrelated Rayleigh-fading samples through simulation with zero mean and unit variance, which are then convoluted with a finite impulse response filter representing the correlation function. Before feeding time-correlated channel data to the ML models, the data is normalized for efficient convergence. Out of the $2550$ time-correlated data samples, \SI{80}{\percent} is allocated for training, while the remaining \SI{20}{\percent} are allocated for validation. Another $50$ samples are taken for testing per Monte Carlo iteration. In all the architectures, a moving window is deployed for iterative training, validation, and testing processes. To learn the temporal patterns, consecutive $10$ samples are taken. Using this learned knowledge, $11^{th}$ sample is predicted. A perfect CSI assumption is made for the training data. Though this is a strong assumption, it helps us to obtain useful information about the performance upper bound.

\section{Results and Discussions}
\label{sec:results}

This section presents comprehensive simulation results to verify the effectiveness of our proposed transformer-based CSI prediction framework. The number of parameters and other architectural details used by different ML strategies are also put into perspective. To demonstrate the performance advantages of our approach, we consider two baseline ML architectures, namely DNN and LSTM, where metrics such as data rate and outage probability are compared.

We consider the path loss exponent for the BS-RIS and the RIS-UE links to be $2.2$, whereas the exponent for the direct BS-UE link is assumed to be $4.2$. The reference path loss value and noise power are taken as \SI{-30}{\decibel} and \SI{-100}{\decibel}, respectively. The distances from the transmitter to the RIS, the RIS to the receiver, and the transmitter to the receiver are set to \SI{38}{m}, \SI{5}{m}, and \SI{40}{m}. Here, the SNR threshold $\gamma_{th} =1$, correspondingly  $R_{\text{th}} =1$ too. Unless stated otherwise, an RIS with eight elements is considered and the transmit power is set to 0dB. 

\subsection{Comparison with ML Baseline Architectures}

In this section, the performance of the proposed transformer approach is compared with the conventional DNN and the LSTM baseline ML architectures. All ML models are trained for $100$ epochs employed with the Adam optimizer alongside the root mean square error (RMSE) as the loss function. ReLU is used as the activation function for DNN and LSTM architectures while Tanh is used for the transformers. Architectures in all the models used a window size of $10$. In each encoder layer, four attention heads and $20$ feedforward layer dimensions are used across all transformer models.

\subsubsection{Variation of Transmit Power with Fixed RIS Elements \vspace{2mm}}

\begin{table*}[!ht]
\centering
\caption{Optimized hyperparameters of DNN, LSTM, and Transformer models.}
\begin{tabular}{|l|*{5}{>{\centering\arraybackslash}p{0.47cm}}*{5}{>{\centering\arraybackslash}p{0.47cm}}*{5}{>{\centering\arraybackslash}p{0.47cm}}|}
\hline
\multirow{2}{*}{\textbf{Hyperparameters}} & \multicolumn{15}{c|}{\textbf{Values}} \\ \cline{2-16} 
 & \multicolumn{5}{c|}{\textbf{DNN}} & \multicolumn{5}{c|}{\textbf{LSTM}} & \multicolumn{5}{c|}{\textbf{Transformer}} \\ \hline
Number of RIS elements & $4$ & $8$ & $12$ & $16$ & \multicolumn{1}{c|}{$20$} & $4$ & $8$ & $12$ & $16$ & \multicolumn{1}{c|}{$20$} & $4$ & $8$ & $12$ & $16$ & $20$ \\
Input and Output features & $16$ & $32$ & $48$ & $64$ & \multicolumn{1}{c|}{$80$} & $16$ & $32$ & $48$ & $64$ & \multicolumn{1}{c|}{$80$} & $16$ & $32$ & $48$ & $64$ & $80$ \\
Neurons in LSTM layer & - & - & - & - & \multicolumn{1}{c|}{-} & $22$ & $44$ & $60$ & $76$ & \multicolumn{1}{c|}{$100$} & - & - & - & - & - \\
Neurons in FC layer 1 & $20$ & $36$ & $52$ & $68$ & \multicolumn{1}{c|}{$88$} & $18$ & $36$ & $56$ & $68$ & \multicolumn{1}{c|}{$90$} & $16$ & $32$ & $48$ & $64$ & $80$ \\
Neurons in FC layer 2 & $24$ & $40$ & $56$ & $72$ & \multicolumn{1}{c|}{$92$} & - & - & - & - & \multicolumn{1}{c|}{-} & - & - & - & - & - \\
Neurons in FC layer 3 & $24$ & $44$ & $56$ & $68$ & \multicolumn{1}{c|}{$94$} & - & - & - & - & \multicolumn{1}{c|}{-} & - & - & - & - & - \\
Neurons in FC layer 4 & $20$ & $36$ & $52$ & $68$ & \multicolumn{1}{c|}{$88$} & - & - & - & - & \multicolumn{1}{c|}{-} & - & - & - & - & - \\
Transformer model dimensions & - & - & - & - & \multicolumn{1}{c|}{-} & - & - & - & - & \multicolumn{1}{c|}{-} & $24$ & $48$ & $60$ & $80$ & $120$ \\ \hline
\end{tabular}
\label{hyperparameters_combined}
\end{table*}

\begin{table}[!ht]
\caption{Summary of metrics associated with DNN, LSTM, and Transformer models.}
\begin{center}

\begin{tabular}{c l c c c c}
\hline
\textbf{RIS}&\textbf{Architecture}&\textbf{Train }&\textbf{Test }& \textbf{\textit{P}} \\
\textbf{Elements}& &\textbf{RMSE}&\textbf{RMSE}& & \\
\hline
 & DNN & $0.0335$ & $0.0271$ & $2,280$ \\
 $4$ & LSTM & $0.0149$ & $0.0172$ & $3,796$ \\
 & Transformer & $0.0134$ & $0.0158$ & $6,838$\\
\hline
 & DNN & $0.0364$ & $0.023$ &  $6,968$ \\
 $8$ & LSTM & $0.0282$ & $0.0191$ & $16,532$ \\
 & Transformer & $0.0131$ & $0.0142$ & $26,702$\\
\hline
 & DNN & $0.0413$ & $0.0259$ & $14,216$ \\
$12$ & LSTM & $0.0329$ & $0.0219$ & $32,552$\\
 & Transformer & $0.0159$ & $0.0157$ & $46,818$\\
\hline
  & DNN & $0.0535$ & $0.0317$ &$24,024$ \\
$16$ & LSTM & $0.0425$ & $0.0266$ & $52,820$ \\
 & Transformer & $0.0177$ & $0.0165$ & $82,894$ \\
\hline
 & DNN & $0.0555$ & $0.0322$ & $39,538$ \\
$20$ & LSTM & $0.036$ & $0.0228$ & $89,170$ \\
 & Transformer & $0.0182$ & $0.0166$ & $164,630$ \\
\hline
\noalign{\vskip 1mm}
\end{tabular}

\label{metrics2}
\end{center}
\end{table}

Let us consider the system model as mentioned in Fig. \ref{fig:system_model} with the number of RIS elements set to $N=8$. There are $16$ channels associated with the RIS model, eight each for the BS-RIS and RIS-UE link, respectively. Since we are predicting real and imaginary channel values separately, $32$ features are required to predict at once. 
Since the number of RIS elements remains constant in this scenario, we can employ a single model for each architecture to make predictions, as the input to each model remains unchanged.

The transmit power is varied from \SI{0}{dBm} to \SI{50}{dBm} by keeping other initial model parameters of the RIS the same. Optimized hyperparameters of the ML models only used for this scenario are tabulated under the columns where the number of RIS elements is eight for each approach as in Table \ref{hyperparameters_combined}. To measure the effectiveness of the architectures, a performance evaluation should be carried out. In this study, we use RMSE as the prediction evaluation criterion. It measures the RMSE between the predicted feature sequence and the actual sequence values in the test and training data sets separately. Apart from that, we have obtained the number of model parameters $P$ associated with models which are calculated using weights and biases. Metrics obtained through the above model simulations are organized in Table \ref{metrics2}.  Moreover, when $N$ is fixed at $8$ it is evident that the proposed transformer-based approach significantly outperforms LSTM and DNN architectures, thereby reducing the training RMSE approximately by \SI{115}{\percent} and \SI{177}{\percent} and test RMSE  approximately by \SI{34}{\percent} and  \SI{62}{\percent}, respectively.

\subsubsection{Variation of RIS Elements with Fixed Transmit Power\vspace{2mm}}

In this scenario, number of RIS elements is the variable. Let us consider instances where the number of RIS elements is $4$, $8$, $12$, $16$, and $20$ by keeping other initial model parameters of the RIS the same. Therefore, the number of channels associated with the RIS are $8, 16, 24, 32$, and $40$, respectively. Then, according to our model configuration (considering real and imaginary values separately as features), the number of features we need to handle would be $16, 32, 48, 64$, and $80$, respectively. For each case, we need to optimize the ML models separately since the number of input features is changing. 
The optimized hyperparameters for each architecture are summarized in Table \ref{hyperparameters_combined}. It shows the number of neurons available in each layer of the respective architecture, the number of neurons in the LSTM layer, and the transformer model dimensions.

Table \ref{metrics2} presents the summary of metrics associated with all three ML architectures such as the train RMSE, the test RMSE, and the number of model parameters. According to the observations, it is clear that the transformer architecture has outperformed the DNN and LSTM architectures in terms of performance when compared with both the train and test RMSE values. 
This shows that the transformer architecture can significantly outperform the state-of-the-art ML architectures, albeit at the cost of higher complexity. Hence, the transformer architecture is preferable in scenarios where the prediction accuracy is to be prioritized over computational complexity. However, the increased computational complexity can be easily handled by using optimized hardware such as tensor processing units.

\subsection{RIS Phase Optimization with Predicted CSI}

\begin{figure*}[ht!]

    \centering
    \hfill
    \begin{minipage}[b]{0.49\textwidth}
        \centering
        \includegraphics[width=\textwidth]{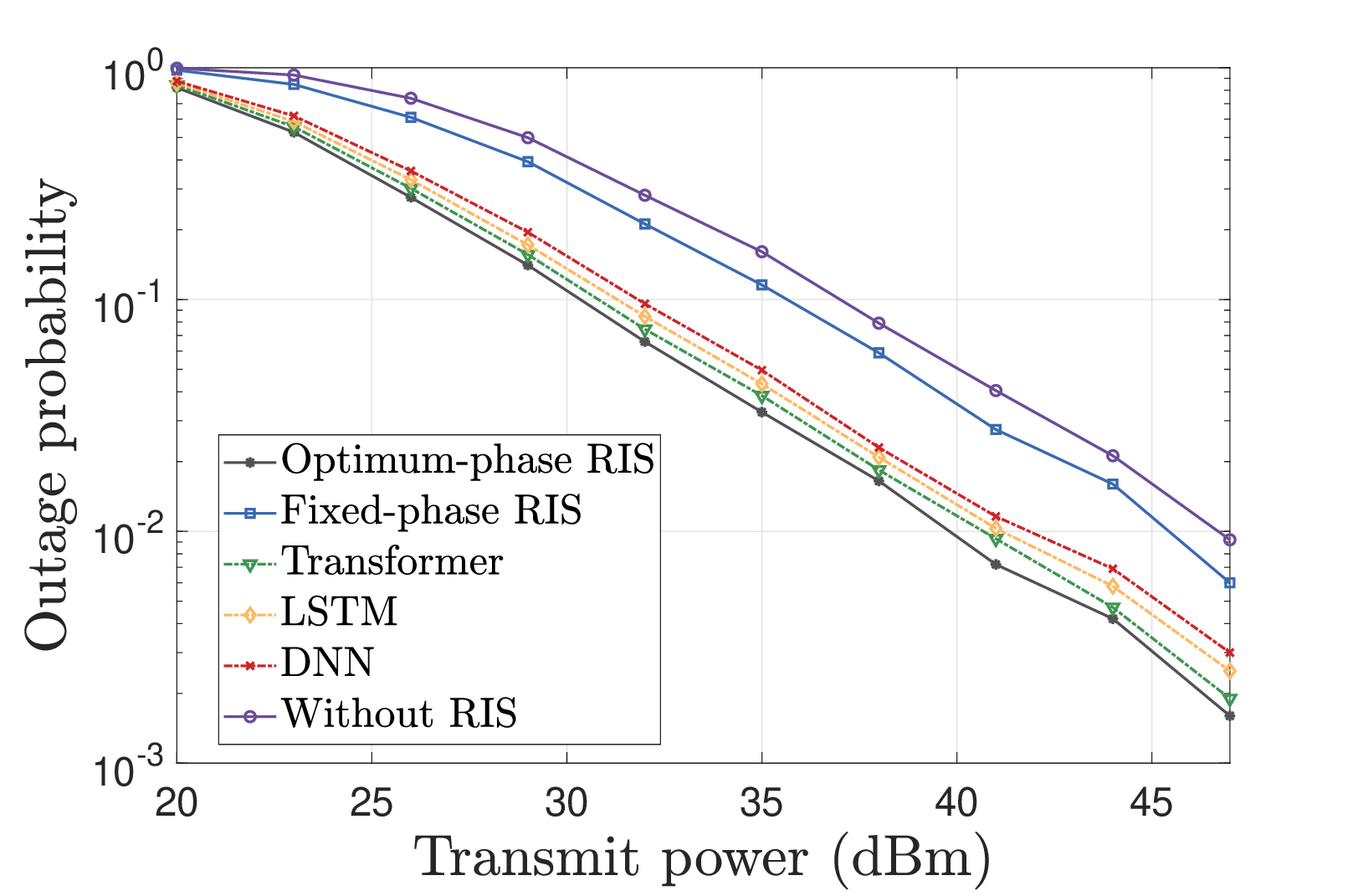}
        \subcaption{}
    \end{minipage}
    \hspace{0.001\linewidth}
    \begin{minipage}[b]{0.49\textwidth}
        \centering
        \includegraphics[width=\textwidth]{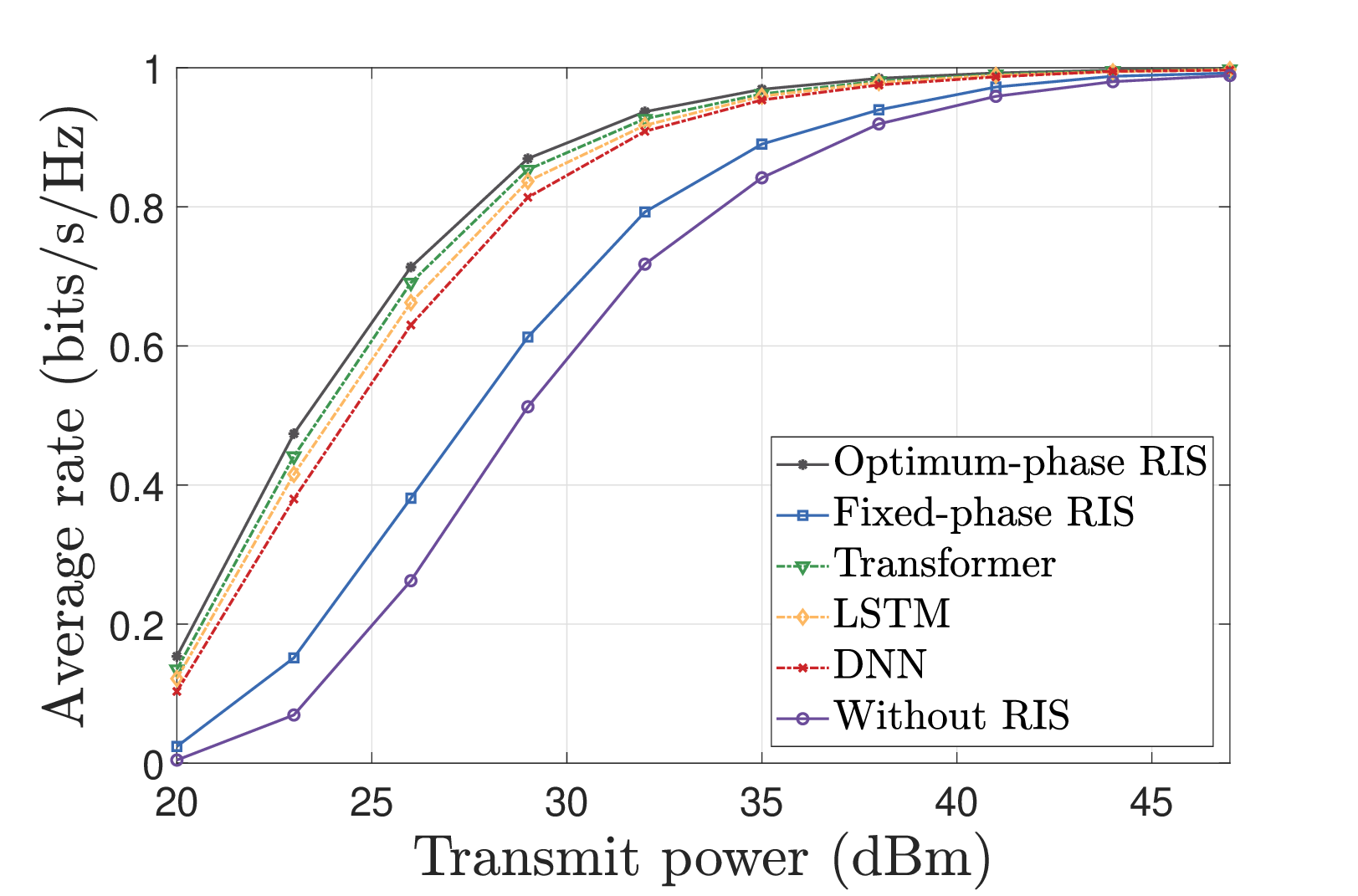}
        \subcaption{}
    \end{minipage}
    \caption{System performance for a fixed number of RIS reflecting elements ($N=8$) in terms of (a)  outage probability, and (b) average rate, when the transmit power is varied.}
    \label{fig:MLTx}
\end{figure*}

\begin{figure*}[ht!]
    \centering
    \begin{minipage}[b]{0.49\textwidth}
        \centering
        \includegraphics[width=\textwidth]{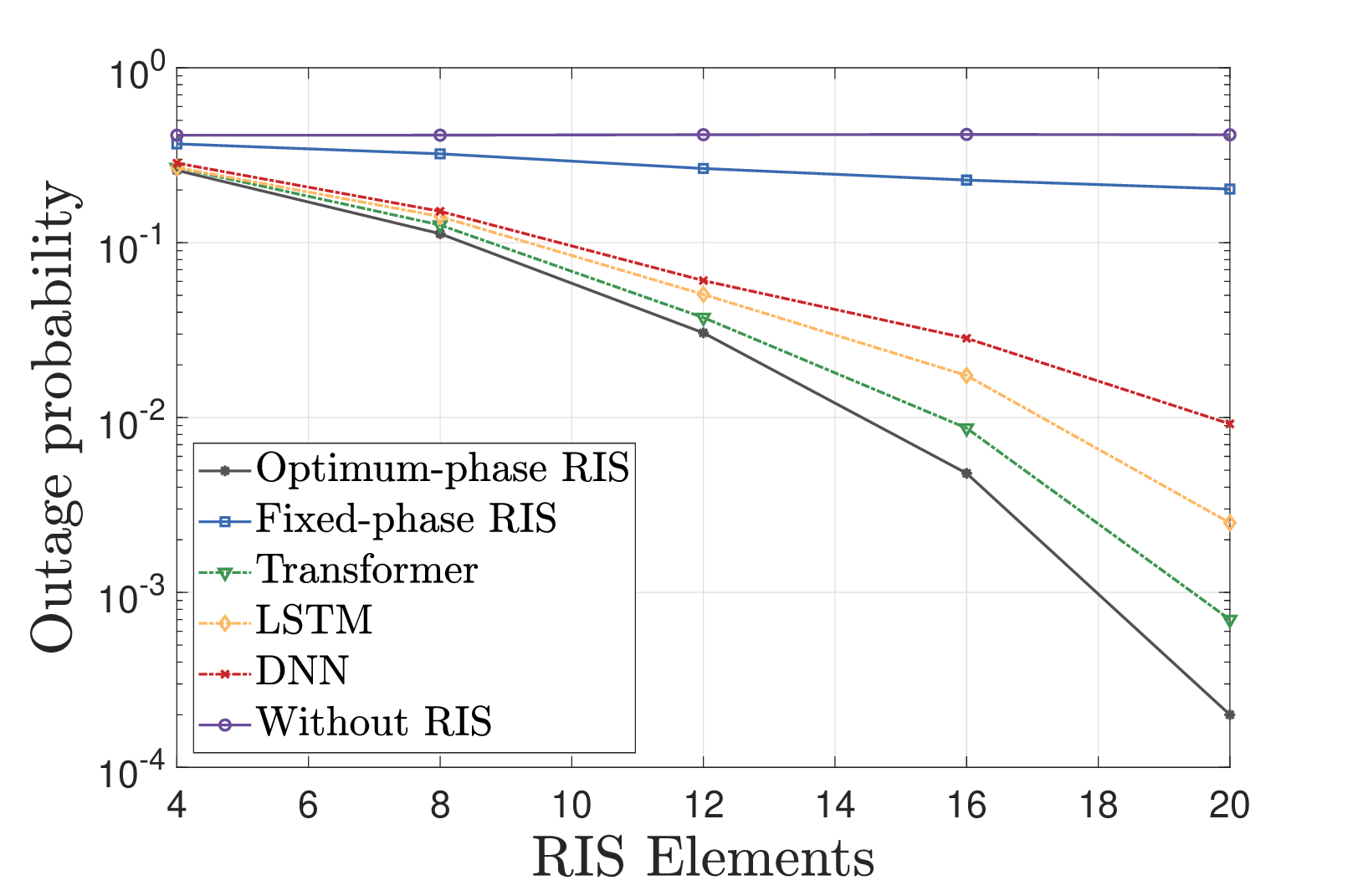}
        \subcaption{}
    \end{minipage}
    \hspace{0.001\linewidth}
    \begin{minipage}[b]{0.49\textwidth}
        \centering
        \includegraphics[width=\textwidth]{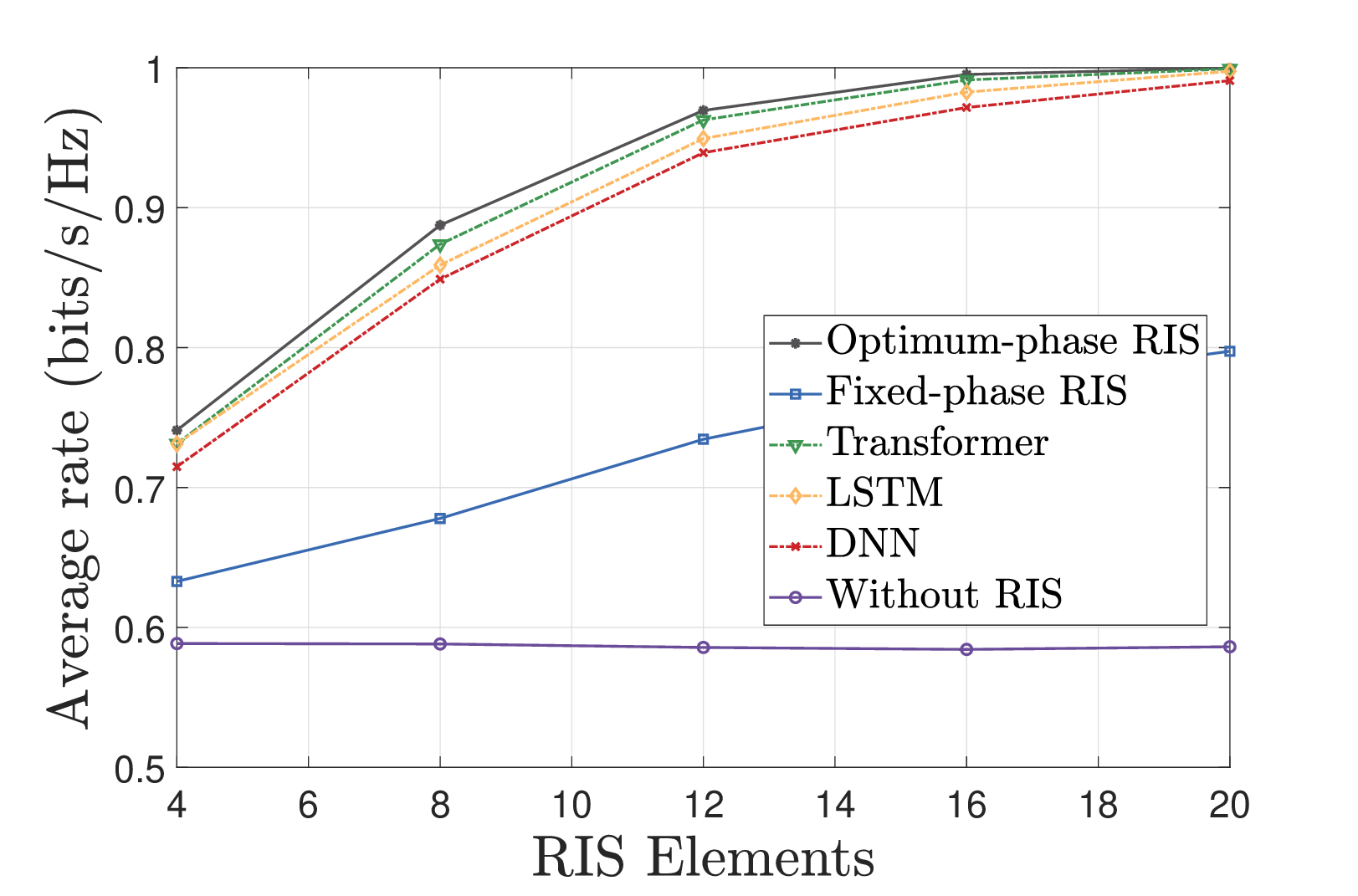}
        \subcaption{}
    \end{minipage}
    \caption{System performance for fixed transmit power (\SI{30}{dBm}) in terms of (a) outage probability, and (b) average rate when the number of RIS reflecting elements is varied.}
    \label{fig:MLN}
\end{figure*}

Using the prediction values obtained from the ML models, we can calculate the optimal phase shift vector of the RIS model as mentioned in \eqref{eqn:phiopt}. Then plugging the test data (actual) channel values and the calculated RIS optimal phase shift vector to \eqref{eqn:gamma}, we can obtain the maximized SNR values. From there onwards, \eqref{eqn:outageprob} and \eqref{eqn:achievable data rate} can be obtained with respect to the maximized SNR values. Therefore, we have shown that RIS could be optimized by predicting the CSI in a RIS-assisted communication system. Let us focus on the two major scenarios of our research output. The generated results shown in Fig. \ref{fig:MLTx} and Fig. \ref{fig:MLN} compare key scenarios, including the outputs with optimized-phase RIS, fixed-phase RIS, without the RIS, and the integrated outputs predicted by the transformer, LSTM, and DNN models. Additionally, in the fixed-phase RIS scenario, $\mathbf{\Phi}$ is set as an identity matrix.

The system performance for a fixed number of RIS elements with varying transmit power is illustrated in Fig. \ref{fig:MLTx}.
Specifically, Figure 4(a) demonstrates the variation in outage probability, while Figure 4(b) shows the variation in average rate. As the transmit power increases, the outage probability decreases, and the average rate increases. In Fig. 4(a), the lowest outage probability is observed when the RIS is optimized. It is the optimum scenario when the CSI is accurately known at the transmitter. This is the performance upper/lower bound and is not possible in practice. The transformer model closely approaches the performance of the optimized RIS, outperforming both the LSTM and DNN models. Furthermore, to achieve an outage probability of $0.01$, the optimized RIS scenario requires a transmit power of \SI{39.79}{dBm}, compared to \SI{46.66}{dBm} for the scenario without RIS. In addition, to achieve the same outage probability, the required transmit powers for the transformer, LSTM, and DNN models are \SI{40.67}{dBm}, \SI{41.13}{dBm}, and \SI{41.85}{dBm}, respectively. In Fig. 4(b), the average rate is maximized when the RIS is optimized and the transformer model gets very close to the upper bound established by the optimized RIS. Moreover, it can be observed that when the transmit power is \SI{25}{dBm}, the performance gap between the scenarios where the transformer model and the system without RIS is approximately \SI{0.4}{bits/s/Hz}. The LSTM and DNN models also demonstrate notable performance improvements, although they fall short of the transformer model's performance.

The system performance for a fixed transmit power, while varying the number of RIS elements, is presented in Fig. \ref{fig:MLN}. 
Figure 5(a) illustrates the variation in outage probability, whereas Figure 5(b) depicts the variation in average rate. Generally, as the number of RIS elements increases, the outage probability decreases, and the average rate increases for scenarios incorporating RIS. Figure 5(a) shows that the outage probability is minimized when the RIS is optimized. The transformer architecture yields the closest results to the optimized RIS, followed by the LSTM and DNN architectures. As per the observations, for a 12-element RIS system, the optimized RIS scenario achieves an outage probability of $0.03$, while the transformer, LSTM, and DNN scenarios result in outage probabilities of $0.037$, $0.05$, and $0.06$, respectively. In Fig. 5(b), the average rate is maximized when the RIS is optimized and the transformer architecture again demonstrates the closest prediction to the average rate achieved with the optimized RIS. When the number of RIS elements reaches eight, the performance gap between the scenario with the transformer model and the scenario without RIS is approximately \SI{0.3}{bits/s/Hz}.

From the results, it is evident that the application of RIS has increased the overall system performance noticeably and the best results are given when the phase is optimized. In every graph, prediction curves lie between the optimum phase and the system without RIS. 
Out of the three prediction curves, the transformer prediction curve goes very closely with the optimum phase RIS output showing that the transformers have provided the most convincing results compared to both the DNN and LSTM architectures.


\section{Conclusions}
\label{sec:conc}
Harnessing the full potential of RIS technology hinges on accurate CSI estimation/prediction since the optimum RIS phase shifts are a function of the corresponding composite channel's CSI. In this study, we proposed a novel sequence-to-one transformer architecture to predict the RIS-associated CSI, enabling the efficient and accurate optimization of the RIS phase shifts. For the CSI prediction task, three architectures namely DNN, LSTM, and transformers were utilized. For the time series channel sample data set created, the transformer architecture provided the lowest RMSE value outperforming DNN and LSTM methods in the scenarios discussed above. After that, the predicted CSI from ML models was fed into the RIS model for phase optimization. According to the results obtained, the transformer was the better multivariate time series prediction architecture out of the three architectures in terms of performance but at the cost of higher complexity. Optimizing the RIS phase shift based on transformer-predicted CSI was found to perform very close to the optimum case when the CSI was assumed to be accurately known. Conversely, the DNN architecture yielded the lowest performance and was also the least complex architecture. The LSTM architecture was positioned in between the other two architectures, offering a middle ground in terms of both performance and complexity. Finally, this study concluded that the proposed sequence-to-one transformer architecture provided promising results for channel prediction regarding RIS phase optimization.


\bibliography{ifacconf}             
                                                   







\end{document}